\documentclass[aps,pra,twocolumn,floatfix,superscriptaddress]{revtex4}
\usepackage{graphicx}
\usepackage{nicefrac}
\usepackage{amsmath}
\usepackage{amsfonts}
\usepackage{amssymb}
\usepackage{amsthm}
\usepackage{epsf}
\usepackage{bm}
\usepackage{bbm}
\usepackage{longtable}

\usepackage{dcolumn}

\newcolumntype{.}{D{x}{}{-1}}

\begin{document}

\newcommand{\half}{\frac12}
\newcommand{\vare}{\varepsilon}
\newcommand{\pr}{^{\prime}}
\newcommand{\ppr}{^{\prime\prime}}
\newcommand{\pp}{{p^{\prime}}}
\newcommand{\hp}{\hat{\bfp}}
\newcommand{\hpp}{\hat{\bfpp}}
\newcommand{\hx}{\hat{\bfx}}
\newcommand{\hq}{\hat{\bfq}}
\newcommand{\hz}{\hat{\bfz}}
\newcommand{\hr}{\hat{\bfr}}
\newcommand{\hn}{\hat{\bfn}}
\newcommand{\rx}{{\rm x}}
\newcommand{\rp}{{\rm p}}
\newcommand{\rmq}{{\rm q}}
\newcommand{\rpp}{{{\rm p}^{\prime}}}
\newcommand{\rk}{{\rm k}}
\newcommand{\bfe}{{\bm e}}
\newcommand{\bfp}{{\bm p}}
\newcommand{\bfpp}{{\bm p}^{\prime}}
\newcommand{\bfq}{{\bm q}}
\newcommand{\bfx}{{\bm x}}
\newcommand{\bfk}{{\bm k}}
\newcommand{\bfz}{{\bm z}}
\newcommand{\bfr}{{\bm r}}
\newcommand{\bfn}{{\bm n}}
\newcommand{\bphi}{{\mbox{\boldmath$\phi$}}}
\newcommand{\bnabla}{\mbox{\boldmath$\nabla$}}
\newcommand{\bgamma}{{\mbox{\boldmath$\gamma$}}}
\newcommand{\balpha}{{\mbox{\boldmath$\alpha$}}}
\newcommand{\bsigma}{{\mbox{\boldmath$\sigma$}}}
\newcommand{\bomega}{{\mbox{\boldmath$\omega$}}}
\newcommand{\bvare}{{\mbox{\boldmath$\varepsilon$}}}

\newcommand{\bGamma}{\bm{\Gamma}}
\newcommand{\intzo}{\int_0^1}
\newcommand{\intinf}{\int^{\infty}_{-\infty}}
\newcommand{\ka}{\kappa_a}
\newcommand{\kb}{\kappa_b}
\newcommand{\ThreeJ}[6]{
        \left(
        \begin{array}{ccc}
        #1  & #2  & #3 \\
        #4  & #5  & #6 \\
        \end{array}
        \right)
        }
\newcommand{\SixJ}[6]{
        \left\{
        \begin{array}{ccc}
        #1  & #2  & #3 \\
        #4  & #5  & #6 \\
        \end{array}
        \right\}
        }
\newcommand{\NineJ}[9]{
        \left\{
        \begin{array}{ccc}
        #1  & #2  & #3 \\
        #4  & #5  & #6 \\
        #7  & #8  & #9 \\
        \end{array}
        \right\}
        }
\newcommand{\Dmatrix}[4]{
        \left(
        \begin{array}{cc}
        #1  & #2   \\
        #3  & #4   \\
        \end{array}
        \right)
        }
\newcommand{\cross}[1]{#1\!\!\!/}
\newcommand{\eps}{\epsilon}
\newcommand{\beq}{\begin{equation}}
\newcommand{\eeq}{\end{equation}}
\newcommand{\beqn}{\begin{eqnarray}}
\newcommand{\eeqn}{\end{eqnarray}}
\newcommand{\lbr}{\langle}
\newcommand{\rbr}{\rangle}
\newcommand{\Za}{Z\alpha}

\title{Two-loop QED corrections with closed fermion loops for the bound-electron $\bm{g}$ factor}

\author{V.~A. Yerokhin}
\affiliation{Max~Planck~Institute for Nuclear Physics, Saupfercheckweg~1, D~69117 Heidelberg,
Germany} \affiliation{ExtreMe Matter Institute EMMI, GSI Helmholtzzentrum f\"ur
Schwerionenforschung, D-64291 Darmstadt, Germany} \affiliation{Center for Advanced Studies,
St.~Petersburg State Polytechnical University, Polytekhnicheskaya 29,
        St.~Petersburg 195251, Russia}

\author{Z. Harman}
\affiliation{Max~Planck~Institute for Nuclear Physics, Saupfercheckweg~1, D~69117 Heidelberg, Germany}
\affiliation{ExtreMe Matter Institute EMMI, GSI Helmholtzzentrum f\"ur
Schwerionenforschung, D-64291 Darmstadt, Germany}

\begin{abstract}

Two-loop QED corrections with closed fermion loops are calculated for the $1s$ bound-electron $g$ factor. Calculations are performed to all orders in the nuclear binding strength parameter $\Za$ (where $Z$ is the nuclear charge and $\alpha$ is the fine structure constant) except for the
closed fermion loop, which is treated within the free-loop (Uehling) approximation in some cases.
Comparison with previous $\Za$-expansion calculations is made and the higher-order remainder of
order $\alpha^2(\Za)^5$ and higher is separated out from the numerical results.

\end{abstract}


\maketitle

Highly charged ions are often considered to be an ideal testing ground for studying bound-state
quantum electrodynamics (QED) effects, in particular, the effects that are non-perturbative in the
binding nuclear strength parameter $\Za$ (where $Z$ is the nuclear charge, $\alpha$ is the fine
structure constant). For light atomic systems, the parameter $\Za$ is small and the $\Za$ expansion
is widely used as a convenient basis for theoretical calculations. However, high accuracy
achieved in modern experiments often demands calculations of QED corrections beyond the $\Za$
expansion even for light atoms. For heavy highly charged ions, the $\Za$ expansion is not
applicable at all and calculations should be only carried out  to all orders in $\Za$.

One of the prominent examples of experiments in light atoms that require for their interpretation
calculations of QED effects to all orders in $\Za$  is the determination of the bound-electron $g$
factor in hydrogenlike ions. A series of spectacular measurements has been accomplished during the
last two decades \cite{haeffner:00:prl,verdu:04,sturm:11,sturm:13:Si,wagner:13}, which brought the
experimental accuracy on the level of few parts in $10^{-11}$. These measurements triggered a large
number of calculations of various QED effects that were required for advancing theory to the level of
experimental interest. In particular, all-order (in $\Za$) calculations of the one-loop self-energy
\cite{yerokhin:02:prl} and nuclear recoil \cite{shabaev:02:recprl} corrections were accomplished,
as well as $\Za$-expansion calculations of the two-loop QED effects
\cite{pachucki:04:prl,pachucki:05:gfact}. The comparison between the experimental and the theoretical
results not only constituted a highly
sensitive test of bound-state QED theory but also led to an accurate determination of
fundamental physical constants such as the electron mass \cite{beier:02:prl,mohr:13:codata}.

Despite all theoretical efforts, the present theory of the bound-electron $g$ factor is not able to
match the experimental accuracy for the heaviest measured ion, Si$^{13+}$ \cite{sturm:13:Si}. The
main reason for this are the two-loop QED effects, which are presently calculated within the $\Za$
expansion up to order $\alpha^2(\Za)^4$ only. The uncertainty due to unknown higher-order two-loop
effects induces the dominant error in the theoretical prediction for ions with $Z>6$. For silicon
with $Z=14$, this uncertainty is already by more than an order of magnitude larger than the
experimental error \cite{sturm:13:Si}. Scaling as $Z^5$, it is going to become even more crucial
for comparison of theory with experiments on heavier-$Z$ ions, which should become feasible in the
near future \cite{sturm:11:prl}.

Calculation of the two-loop QED corrections to all orders in the nuclear binding strength parameter
$\Za$ is a very difficult task. Such calculation for the Lamb shift in hydrogenlike ions extended
for over a decade (see Refs.~\cite{yerokhin:08:twoloop,yerokhin:09:sese,yerokhin:10:sese} for the
present status). A similar calculation for the bound-electron $g$ factor should be feasible in
principle but is going to be even more difficult than for the Lamb shift, for several reasons.
First, Feynman diagrams for the $g$ factor contain an additional vertex representing the
interaction with the external magnetic field as compared to the diagrams contributing to the Lamb
shift. Second, the convergence of the partial-wave expansion (which is usually the limiting factor
for the accuracy of calculations) is typically slower for the $g$ factor than for the Lamb shift.
Third, the unknown higher-order remainder to the Lamb shift is suppressed by the factor of
$(\Za)^2$ with respect to the leading contribution, whereas for the $g$ factor the suppression
factor is $(\Za)^5$.

The two-loop QED effects can be separated into two large pieces, the two-loop self-energy
correction and the two-loop corrections with closed fermion (vacuum-polarization) loops. In the
present study, we consider the latter part, leaving the two-loop self-energy (being the most
nontrivial part) for future investigations. Calculations of the two-loop corrections with
vacuum-polarization loops are simplified by the fact that such loops can be treated within the
free-loop (Uehling) approximation, which replaces the loop of the bound-electron propagators by the
leading term of its expansion in the binding potential. In the one-loop case, such approximation
leads to the well-known Uehling potential and induces the dominant part of the one-loop
vacuum-polarization effect even for ions as heavy as uranium. In the present investigation, we
employ the free-loop approximation for some corrections, namely the self-energy correction with the
vacuum-polarization insertion into the photon line and the two-loop vacuum-polarization correction.
In addition, there are several diagrams that vanish in the free-loop approximation, namely the
diagrams with the interaction with the external magnetic field attached to the vacuum-polarization
loop. The contribution of such diagrams should be small, so they are omitted in the present
investigation.

The remaining paper is organized as follows. In the next three sections, we study three
gauge-invariant subsets of two-loop contributions with vacuum-polarization loops. Namely, the
self-energy correction with the vacuum-polarization insertion into the photon line is calculated in
Sec.~\ref{sec:svpe}, the self-energy correction with the vacuum-polarization insertion into the
electron line is calculated in Sec.~\ref{sec:sevp}, and the two-loop vacuum-polarization correction
is calculated in Sec.~\ref{sec:vpvp}. In the last section, we summarize the results obtained and
discuss the experimental consequences of our calculations.

The relativistic units ($\hbar=c=1$) are used in this paper. We will also use the abbreviations
"SE" for the self-energy and "VP" for the vacuum-polarization.


\section{
Self-energy correction with vacuum-polarization insertion into the photon line}

\label{sec:svpe}

\begin{figure}
\centerline{
\resizebox{0.5\textwidth}{!}{%
  \includegraphics{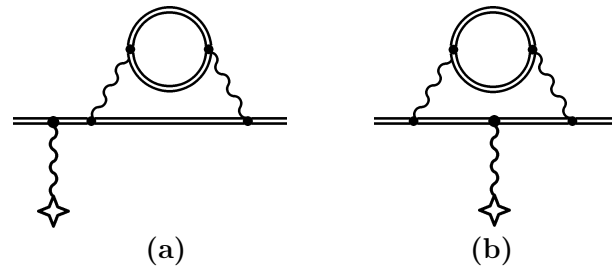}
}}
 \caption{
 The
combined self-energy and vacuum-polarization correction to the bound-electron $g$ factor with
vacuum-polarization insertion into the photon line, referred to as the S(VP)E correction. Double
lines represent an electron propagating in the binding nuclear field, wave lines denote virtual
photon, and the wave line terminated by a cross denotes interaction with an external magnetic
field.
 \label{fig:svpe}}
\end{figure}

We start with the set of combined SE-and-VP diagrams depicted on Fig.~\ref{fig:svpe}, whose
contribution will be referred to as the S(VP)E correction. This correction can be regarded as the
one-loop SE correction to the $g$ factor in which the standard photon line in the SE loop is
substituted by the ``dressed'' photon line with the VP insertion. In this section, we will treat
the VP insertion within the free-loop approximation only. The part of the S(VP)E diagram beyond
this approximation involves a light-by-light scattering subdiagram, whose calculation is
notoriously difficult but which usually leads to small effects.

The dressed photon propagator with the free-loop VP insertion can be derived \cite{bogoljubov:book}
in the form of an extension of the standard photon propagator, both in momentum and coordinate
space. In the momentum space with $D = 4-2\epsilon$ dimensions, the dressed VP photon propagator is
given by \cite{mallampalli:96,adkins:98}
\begin{align} \label{b01}
D_{\rm VP}^{\mu\nu}(k) = \frac{\alpha}{\pi}\, I_{\rm VP}(k)\,D^{\mu\nu}(k)\,,
\end{align}
where $D^{\mu\nu}(k)$ is the standard photon propagator,
\begin{align} \label{b02}
I_{\rm VP}(k) =  -C_{\eps}\, k^2
      \int_0^1dz\,
        \frac{z^2(1-z^2/3)}{4\, [m^2-k^2(1-z^2)/4-i0]^{1+\eps}}\,,
\end{align}
$C_{\eps} = (4\pi)^{\eps}\,\Gamma(1+\eps)$ and $k$ is a four-vector with $k^2 = k_0^2-\bm{k}^2$.
In the coordinate space, the expression for the dressed VP photon propagator reads
\begin{align}\label{b09}
    D^{\mu\nu}_{\rm VP}(\omega,\bfx_{12}) = \frac{\alpha}{\pi}\,
      \int_1^{\infty}dt\, I_{\rm VP}(t)\,
       D^{\mu\nu}(\omega,\bfx_{12};2mt)\,,
\end{align}
where $D^{\mu\nu}(\omega,\bfx;2mt)$ is the standard propagator of a massive photon with mass
$\lambda = 2mt$ and
\begin{align}\label{b09a}
I_{\rm VP}(t) =  \sqrt{t^2-1}\, \frac{2t^2+1}{3t^4}\,.
\end{align}
In the Feynman gauge, the standard propagator of a massive photon is given by
\begin{equation}\label{b010}
 D^{\mu\nu}(\omega,{\bfx}_{12};\lambda) = g^{\mu\nu}\,
   \frac{\exp[i\sqrt{\omega^2-\lambda^2+i0}\,x_{12}]}{4\pi\,x_{12}}\,.
\end{equation}
The above formulas demonstrate that the dressed VP photon propagator can be effectively obtained
from the standard propagator of a massive photon (multiplied by a simple function) by integrating
over the effective photon mass. Employing this fact, we can construct the calculation of the S(VP)E
correction to the $g$ factor as an extension of our previous calculations of the one-loop SE
correction to the $g$ factor \cite{yerokhin:02:prl,yerokhin:04} and the S(VP)E correction to the
Lamb shift \cite{yerokhin:08:pra}.

Following the approach described in details in Ref.~\cite{yerokhin:04}, we represent
the S(VP)E correction to the $g$ factor as a sum of three contributions,
\begin{equation}\label{b011}
\Delta g_{\rm SVPE} = \Delta g_{\rm ir} + \Delta g_{\rm vr}^{(0)}
 + \Delta g_{\rm vr}^{(1+)}\,.
\end{equation}
The first term on the right-hand-side of the above equation, $\Delta g_{\rm ir}$, is the
irreducible contribution, which is induced by the irreducible ($n\neq a$) part of the diagram on
Fig.~\ref{fig:svpe}a. The second term $\Delta g_{\rm vr}^{(0)}$ is the contribution of the
free-electron propagators in the vertex part (induced by the diagram on Fig.~\ref{fig:svpe}b) and
the reducible part (induced by the reducible $n = a$ part of the diagram on Fig.~\ref{fig:svpe}a).
The third term $\Delta g_{\rm vr}^{(1+)}$ is the remainder of the vertex and reducible parts that
contains one or more interactions with the nuclear biding field in the electron propagators.

The irreducible part $\Delta g_{\rm ir}$ is relatively straightforward to calculate. It can be
represented by a non-diagonal matrix element of the operator responsible for the S(VP)E correction
to the Lamb shift. So, we calculate $\Delta g_{\rm ir}$ by generalizing the method developed by us
for the calculation of the S(VP)E correction to the energy levels \cite{yerokhin:08:pra}.

The zero-potential contribution $\Delta g_{\rm vr}^{(0)}$ is calculated similarly to the
corresponding contribution to the SE correction to the $g$ factor from Ref.~\cite{yerokhin:04}.
Some additional care is required in this case, however, as the free SE and vertex operators with
the VP insertion are more complicated and, in particular, possess a higher degree of UV divergence
than the corresponding one-loop operators ($\sim 1/\epsilon^2$ versus $\sim 1/\epsilon$).
Evaluation of the operators in momentum space and final calculational formulas are summarized in
Appendix~\ref{app:1}.

Numerical results of our calculations of the S(VP)E correction for the $1s$ bound-state $g$ factor
are presented in Table~\ref{tab:svpe}. The calculation was performed for the point nuclear charge.
The uncertainty quoted in the table originates predominantly from the truncation of the
partial-wave expansion in the many-potential vertex and reducible contributions. In our
calculations, we included about 40 partial waves and extrapolated the expansion to infinity by
least-squares fitting the partial sums to a polynomial in inverse cutoff parameter.

In order to improve convergence of the partial-wave expansion and better to estimate the accuracy
of our extrapolation, we employed a modification of the standard potential-expansion
renormalization approach first suggested in Ref.~\cite{yerokhin:01:hfs}. In this modified approach,
the energy in the zero-potential contribution is shifted from its physical value $\vare_a$,
$\vare_a \to \widetilde{\vare}_a = (\vare_a+m)/2$, where $m$ is the electron rest mass. The effect
of this shift is compensated in the many-potential term, which is evaluated as a point-by-point
difference of the unrenormalized contribution and the free-propagator contribution (with exactly
the same energy $\widetilde{\vare}_a$ as in the zero-potential term). As a result, the sum of the
zero-potential and many-potential contributions should not depend on the particular choice of
$\widetilde{\vare}_a$. Individual terms of the partial-wave expansion, however, depend strongly on
$\widetilde{\vare}_a$. Comparing the final results for the sum of the zero- and many-potential
terms for different choices of the free parameter $\widetilde{\vare}_a$, we were able to
cross-check our estimation of uncertainty of the extrapolation of the partial-wave expansion.

\begin{figure*}
\centerline{
\resizebox{0.8\textwidth}{!}{%
  \includegraphics{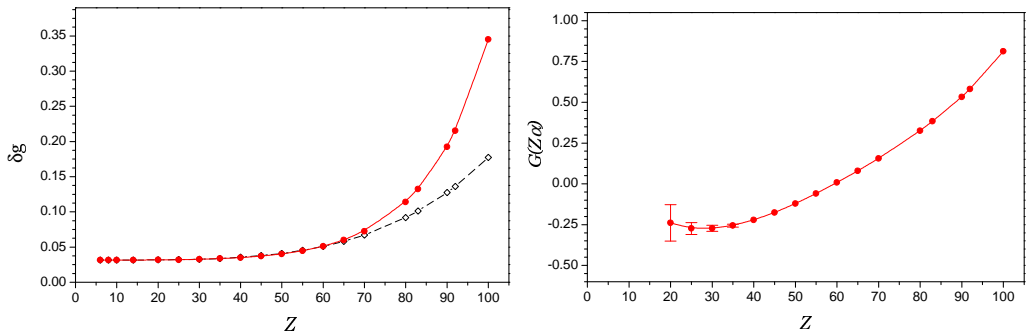}
}}
 \caption{(Color online) The S(VP)E correction to the $1s$ bound-electron $g$ factor. Left graph presents our
numerical all-order results (dots and solid line, red), together with the $\Za$-expansion result
(dashed line, black), in terms of $\delta g = \Delta g/(\alpha^2/\pi^2)$. Right graph
shows the corresponding results for the higher-order remainder $G(\Za)$.
 \label{fig:svpe:res}}
\end{figure*}

In Fig.~\ref{fig:svpe:res}, our numerical data are compared with the results obtained previously
within the $\Za$-expansion approach. The $\Za$ expansion of the S(VP)E correction reads
\begin{align}
\Delta g_{\rm SVPE} &\ = \left(\frac{\alpha}{\pi}\right)^2\, \biggl[
a_0 + (\Za)^2 \,a_2 + (\Za)^4\,a_4
\nonumber \\ &
+ (\Za)^5\,G_{\rm SVPE}(\Za) \biggr]\,,
\end{align}
where the expansion coefficients $a_i$ for the $1s$ state are given by \cite{pachucki:05:gfact}
\begin{align}
a_0 &\ = 0.031\,374\,844\,,\\
a_2 &\ = a_0/6\,,\\
a_4 &\ = 0.504\,539\,572\,,
\end{align}
and $G_{\rm SVPE}(\Za)$ is the higher-order remainder.

The results summarized in Table~\ref{tab:svpe} indicate that the irreducible term $\Delta g_{\rm
ir}$ induces a negligible contribution to the total correction in the low-$Z$ region, whereas for
high $Z$ it is clearly the dominant contribution. The low-$Z$ behavior of this term agrees with the
fact that the first two terms of the $\Za$ expansion of the S(VP)E correction originate from the
anomalous magnetic moment of the electron (i.e., from the vertex contribution) only.

We observe remarkable agreement between our numerical and the $\Za$-expansion results. Noticeable
difference arises only for heavy ions with $Z > 70$. This is due to a combination of factors that
the higher-order remainder $G(\Za)$ (i) is highly suppressed [by a factor of $(\Za)^5$], (ii) is
small numerically, and (iii) changes its sign around $Z = 60$.

It can be seen that the accuracy of our numerical results is not high enough to  directly identify
the higher-order remainder $G(\Za)$ for light ions with $Z < 20$. In order to get $G(\Za)$ for the
experimentally interesting cases of carbon, oxygen, silicon, and calcium, we extrapolated our
results towards lower values of $Z$. For this, we used the extrapolation procedure suggested in
Ref.~\cite{mohr:75:prl}. The results of such extrapolation are: $G_{\rm extr}(Z = 20) =
-0.230(21)$, $G_{\rm extr}(Z = 14) = -0.152(43)$, $G_{\rm extr}(Z = 8) = -0.05(12)$, and $G_{\rm
extr}(Z = 6) = -0.00(15)$.

\begin{table}
\setlength{\LTcapwidth}{\columnwidth} \caption{The S(VP)E correction to the $1s$ bound-electron $g$
factor, in terms of $\delta g = \Delta g/(\alpha^2/\pi^2)$. The second, third and forth columns
summarize the results of the present calculations, whereas the last column presents results
obtained within the $\Za$-expansion approach.
 \label{tab:svpe} }
\begin{ruledtabular}
\begin{tabular}{l....}
                $Z$
                & \multicolumn{1}{c}{ir}
                & \multicolumn{1}{c}{vr}
                & \multicolumn{1}{c}{Total}
                                         & \multicolumn{1}{c}{$Z\alpha$-expansion} \\
\hline\\[-9pt]
  6 &    0.000x\,07 &    0.031x\,32 &    0.031x\,39 &    0.031x\,39 \\
  8 &    0.000x\,12 &    0.031x\,28 &    0.031x\,40(1) &    0.031x\,40 \\
 10 &    0.000x\,20 &    0.031x\,22 &    0.031x\,42(1) &    0.031x\,42 \\
 14 &    0.000x\,41 &    0.031x\,07 &    0.031x\,48(1) &    0.031x\,48 \\
 20 &    0.000x\,96 &    0.030x\,74 &    0.031x\,70(1) &    0.031x\,72 \\
 25 &    0.001x\,68 &    0.030x\,38 &    0.032x\,05(1) &    0.032x\,11 \\
 30 &    0.002x\,72 &    0.029x\,92 &    0.032x\,65(1) &    0.032x\,78 \\
 35 &    0.004x\,21 &    0.029x\,38 &    0.033x\,58(1) &    0.033x\,86 \\
 40 &    0.006x\,27 &    0.028x\,74 &    0.035x\,01(1) &    0.035x\,48 \\
 45 &    0.009x\,12 &    0.028x\,01 &    0.037x\,14(1) &    0.037x\,81 \\
 50 &    0.013x\,03 &    0.027x\,20 &    0.040x\,23(1) &    0.041x\,01 \\
 55 &    0.018x\,39 &    0.026x\,30 &    0.044x\,70(1) &    0.045x\,31 \\
 60 &    0.025x\,71 &    0.025x\,34 &    0.051x\,05(1) &    0.050x\,92 \\
 65 &    0.035x\,69 &    0.024x\,32 &    0.060x\,01(1) &    0.058x\,09 \\
 70 &    0.049x\,24 &    0.023x\,28 &    0.072x\,52(1) &    0.067x\,09 \\
 80 &    0.092x\,58 &    0.021x\,27 &    0.113x\,85(1) &    0.091x\,76 \\
 83 &    0.111x\,64 &    0.020x\,75 &    0.132x\,39(1) &    0.101x\,19 \\
 90 &    0.172x\,67 &    0.019x\,87 &    0.192x\,55(1) &    0.127x\,50 \\
 92 &    0.195x\,67 &    0.019x\,75 &    0.215x\,42(2) &    0.136x\,23 \\
100 &    0.324x\,95 &    0.020x\,21 &    0.345x\,16(2) &    0.177x\,23 \\
\end{tabular}
\end{ruledtabular}
\end{table}

%
%
\section{
Self-energy correction with vacuum-polarization insertion into the electron line}

\label{sec:sevp}

\begin{figure*}
\centerline{\includegraphics[width=\textwidth]{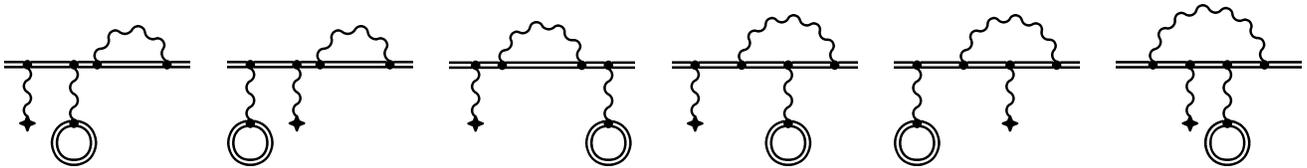}} \caption{The combined self-energy
and vacuum-polarization correction to the bound-electron $g$ factor with vacuum-polarization
insertion into the electron line, referred to as the SEVP correction.
 \label{fig:sevp} }
\end{figure*}

We now turn to the set of combined SE-and-VP diagrams depicted on Fig.~\ref{fig:sevp}, whose
contribution will be referred to as the SEVP correction. This correction can be regarded as the
one-loop SE correction to the $g$ factor, in which one of the bound-electron propagators is
modified by the VP insertion. Since the VP insertion into the electron line can be represented by a
local potential, the simplest way to calculate the SEVP correction is to redefine the
bound-electron propagator by adding the VP potential to the binding nuclear potential. In this
case, the SEVP correction can be obtained as a difference of the $g$ factor values with and without
the VP addition in the binding potential.

The dominant part of the one-loop VP potential is given by the well-known Uehling potential:
\begin {eqnarray} \label{uehlexpr}
V_{\rm Uehl}(r)&=&-\Za \frac{2\alpha}{3\pi} \int\limits_0^\infty dr'\; 4\pi r'\rho (r') \nonumber
\\ &&\times \int\limits_1^\infty dt \; \left(1 +\frac{1}{2t^2}\right)
\frac{\sqrt{t^2-1}}{t^{2}}\nonumber \\
&&\times \frac{e^{-2m|r-r'|t}-e^{-2m(r+r')t}}
{4mrt} \,,\nonumber \\
\end{eqnarray}
where
$Z\rho(r)$ is the density of the nuclear charge distribution ($\int \rho(r) d{\bf r}=1$). The
remaining part of the one-loop VP potential is given by the so-called Wichmann-Kroll
potential $V_{\rm WK}$. For
the purpose of the present investigation, it is sufficient to evaluate it by approximate
formulas obtained in Ref.~\cite{fainshtein:91}. The one-loop VP potential is then obtained as a sum
of the Uehling and Wichmann-Kroll parts, $V_{\rm VP}(r) = V_{\rm Ueh}(r) + V_{\rm WK}(r)$.

In the present work, we calculate the SEVP correction by calculating the SE correction to the $g$
factor in the combined Coulomb and VP binding potential and subtracting the corresponding
contribution evaluated with the Coulomb potential only. The result obtained in this way contains
small additional contributions induced by second and higher-order iterations of the VP potential,
but they may be disregarded at the present level of interest. The general scheme of calculation of
the SE correction to the bound-electron $g$ factor was developed in our previous studies
\cite{yerokhin:02:prl,yerokhin:04}. In the present work, we extended this scheme for the case of
the general binding potential. To this end, we employed the numerical approach for evaluation of
the Dirac Green function for the arbitrary spherically symmetric potential (behaving as $\sim 1/r$
for $r\to \infty$) developed in Ref.~\cite{yerokhin:11:fns}.

Numerical results of our calculations of the SEVP correction for the $1s$ bound-state $g$ factor
are presented in Table~\ref{tab:sevp}. Our calculation was performed with the Fermi model of the
nuclear charge distribution. The one-loop VP potential included both the Uehling and the
Wichmann-Kroll contributions.

\begin{figure*}
\centerline{\includegraphics[width=0.8\textwidth]{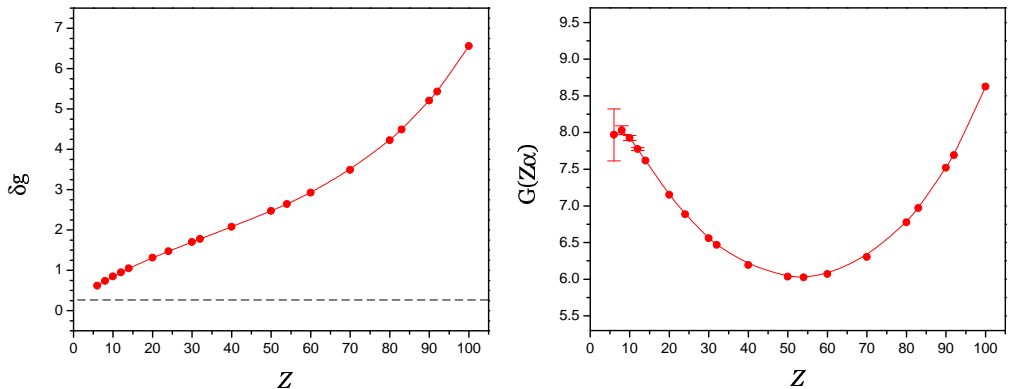}} \caption{(Color online) SEVP
correction to the $1s$ bound-electron $g$ factor. Left graph presents our numerical all-order
results (dots and solid line, red), together with the $\Za$-expansion result (dashed line, black),
in terms of $\delta g = \Delta g/[(\alpha^2/\pi^2)(\Za)^4]$. Right graph shows the results for the
higher-order remainder $G(\Za)$.
 \label{fig:sevp:res} }
\end{figure*}

Comparison of our numerical results with the $\Za$-expansion results is shown graphically on
Fig.~\ref{fig:sevp:res}. The $\Za$ expansion of the SEVP correction is given by
\cite{pachucki:05:gfact}
\begin{align}
\Delta g_{\rm SEVP} &\ = \left(\frac{\alpha}{\pi}\right)^2\, (\Za)^4\,\biggl[ \frac4{15} +
(\Za)\,G_{\rm SEVP}(\Za) \biggr]\,,
\end{align}
where  $G_{\rm SEVP}(\Za)$ is the higher-order remainder. Note that the $\Za$ expansion of the SEVP
correction starts with the $(\Za)^4$ term, so that the higher-order remainder is suppressed only by
first power of $\Za$ in this case.

The results of our calculations indicate that the higher-order contribution for the SEVP correction
is remarkably large. Even for light systems like carbon and oxygen, the total correction is twice
as large as the leading-order contribution, which is rather unusual. Notably, a large higher-order
contribution stemming from the SEVP correction was previously reported also for the Lamb shift
\cite{yerokhin:08:twoloop}.

\begin{table}
\setlength{\LTcapwidth}{\columnwidth} \caption{The SEVP correction to the $1s$ bound-electron $g$
factor, in terms of $\delta^{(4)} g = \Delta g/[(\alpha^2/\pi^2)(\Za)^4]$.
 \label{tab:sevp} }
\begin{ruledtabular}
\begin{tabular}{l..}
                $Z$
                & \multicolumn{1}{c}{$\delta^{(4)} g$}
                & \multicolumn{1}{c}{$\Za$-expansion}
 \\
\hline\\[-9pt]
  6 &  0.6x2(2)     &   0.2x667 \\
  8 &  0.7x36(3)    &   0.2x667 \\
 10 &  0.8x45(2)    &   0.2x667 \\
 12 &  0.9x48(2)    &   0.2x667 \\
 14 &  1.0x45(1)    &   0.2x667 \\
 20 &  1.3x10\,3(9)   &   0.2x667 \\
 24 &  1.4x72\,5(8)   &   0.2x667 \\
 30 &  1.7x02\,2(8)   &   0.2x667 \\
 32 &  1.7x76\,7(4)   &   0.2x667 \\
 40 &  2.0x74\,1(4)   &   0.2x667 \\
 50 &  2.4x67\,8(2)   &   0.2x667 \\
 54 &  2.6x40\,3(3)   &   0.2x667 \\
 60 &  2.9x24\,3(2)   &   0.2x667 \\
 70 &  3.4x86\,0(6)   &   0.2x667 \\
 80 &  4.2x22\,3(6)   &   0.2x667 \\
 83 &  4.4x88\,0(3)   &   0.2x667 \\
 90 &  5.2x06\,1(7)   &   0.2x667 \\
 92 &  5.4x29\,6(9)   &   0.2x667 \\
100 &  6.5x61(2)    &   0.2x667 \\
\end{tabular}
\end{ruledtabular}
\end{table}

%
%
\section{
Two-loop vacuum-polarization correction}

\label{sec:vpvp}

\begin{figure*}
\centerline{\includegraphics[width=0.8\textwidth]{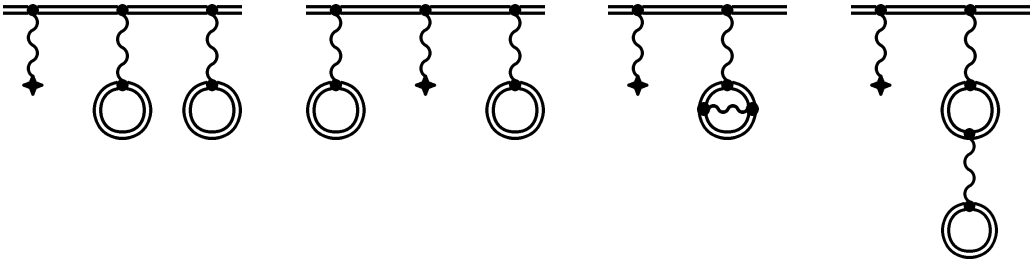}}
\caption{The two-loop vacuum-polarization correction to the bound-electron
$g$ factor (the VPVP
correction).
 \label{fig:vpvp} }
\end{figure*}

In this section we calculate the set of two-loop VP diagrams depicted in Fig.~\ref{fig:vpvp},
referred to as the VPVP correction. The four diagrams in the set can be divided into two parts. The
two diagrams on the left are induced by the second-order iteration of the one-loop VP potential,
whereas the two diagrams on the right are induced by the two-loop VP potential. The complete form
of the two-loop VP potential is not known at present. Its dominant part, however, is delivered by
the free-loop approximation and is known for a long time, first derived by K\"all\'{e}n and Sabry
\cite{kaellen:55}. In the present work, we also treat the two-loop VP potential within the
free-loop approximation only.

The VPVP contribution is given by the following expression
\begin{align}
\Delta g_{\rm VPVP} = &\  2\,\lbr \delta_{\rm VP}a|V_{\rm VP}|\delta_g a\rbr
   - 2\,\lbr a|V_{\rm VP}|a\rbr\, \lbr \delta_{\rm VP} a|\delta_{g} a\rbr
   \nonumber \\ &
   + \lbr \delta_{\rm VP}a|V_{g}|\delta_{\rm VP} a\rbr
   - \lbr a|V_{g}|a\rbr\, \lbr \delta_{\rm VP} a|\delta_{\rm VP} a\rbr
   \nonumber \\ &
   + 2\, \lbr a|V_{\rm KS}|\delta_g a\rbr\,,
\end{align}
where $|a\rbr$ is the reference-state wave function with a fixed momentum projection $\mu = 1/2$,
$|\delta_{\rm VP} a\rbr$ and $|\delta_{g}
a\rbr$ are first-order perturbations of the reference-state wave function by the one-loop
VP potential $V_{\rm VP}$ and the effective $g$-factor potential $V_g$,
\begin{align}
|\delta_{\rm VP} a\rbr = \sum_{n\neq a} \frac{|n\rbr\lbr n|V_{\rm VP}|a\rbr}{\vare_a-\vare_n}\,,
\end{align}
\begin{align}
|\delta_{g} a\rbr = \sum_{n\neq a} \frac{|n\rbr\lbr n|V_{g}|a\rbr}{\vare_a-\vare_n}\,,
\end{align}
the effective $g$-factor potential $V_g$ is
\begin{align}
V_g = 2m\,[\bm{r}\times\bm{\alpha}]_z\,,
\end{align}
and $V_{\rm KS}$ is the K\"all\'{e}n-Sabry potential (see, e.g., Ref.~\cite{fullerton:76} for explicit formulas).
Note that the effective potential $V_g$ is defined so that its matrix element on the reference-state wave function with momentum projection
$\mu = 1/2$ is the Dirac value of the bound-electron $g$ factor, $\lbr a|V_{g}|a\rbr = g_D$.

\begin{figure*}
\centerline{\includegraphics[width=0.8\textwidth]{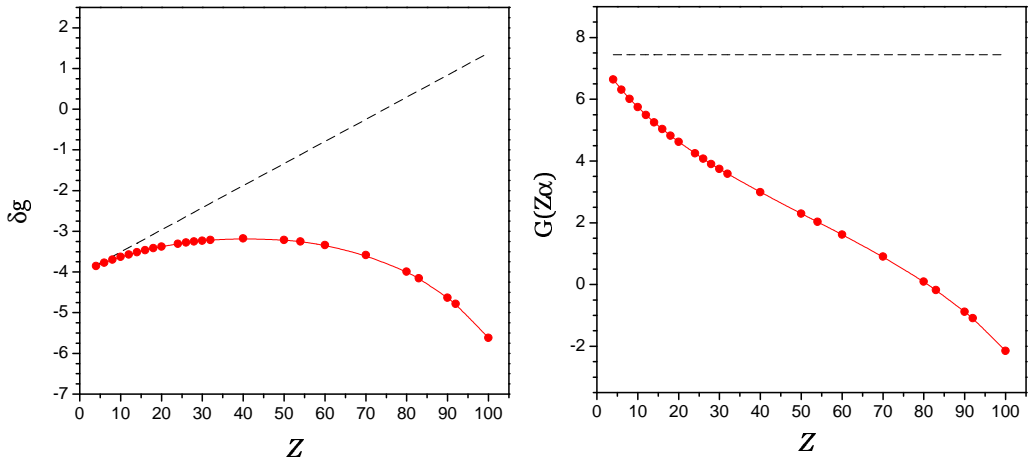}}
\caption{(Color online) VPVP correction to the $1s$ bound-electron $g$ factor. Left graph presents our
numerical all-order results (dots and solid line, red), together with the $\Za$-expansion results
(dashed line, black), in terms of $\delta g = \Delta g/[(\alpha^2/\pi^2)(\Za)^4]$. Right graph
shows the corresponding results for the higher-order remainder $G(\Za)$.
 \label{fig:vpvp:res} }
\end{figure*}

The numerical calculation of the VPVP correction is quite straightforward. It was performed by
obtaining the perturbed wave functions with help of the dually kinetically balanced $B$-spline
basis set method \cite{shabaev:04:DKB}. Numerical results for the VPVP correction for the $1s$
bound-electron $g$ factor are presented in Table~\ref{tab:vpvp}. Comparison of our numerical
all-order results with the $\Za$ expansion results is given in Table~\ref{tab:vpvp} and graphically
in Fig.~\ref{fig:vpvp:res}. The $\Za$ expansion of the VPVP correction reads
\begin{align}
\Delta g_{\rm VPVP} &\ = \left(\frac{\alpha}{\pi}\right)^2\, (\Za)^4\,\biggl[-\frac{328}{81} +
(\Za)\,G_{\rm VPVP}(\Za) \biggr]\,,
\end{align}
where $G_{\rm VPVP}(\Za)$ is the higher-order remainder. The leading term of its $\Za$ expansion
was obtained previously in Ref.~\cite{jentschura:09}, $G_{\rm VPVP}(\Za) = 7.4415187 + O(\Za)$.

We observe that our numerical all-order results agree well with the previous $\Za$-expansion
results. In particular, they confirm the conclusion of Ref.~\cite{jentschura:09} that the
higher-order VPVP contribution $G(\Za)$ is rather large in the low-$Z$ region. In the high-$Z$
region, however, the large contribution of the $(\Za)^5$ coefficient is compensated by higher-order
terms, so that the total value of $G(\Za)$ is significantly reduced and even changes its sign
eventually as $Z$ increases.

\begin{table}
\setlength{\LTcapwidth}{\columnwidth} \caption{VPVP correction to the $1s$ bound-electron $g$
factor, in terms of $\delta^{(4)} g = \Delta g/[(\alpha^2/\pi^2)(\Za)^4]$.
 \label{tab:vpvp} }
\begin{ruledtabular}
\begin{tabular}{l..}
                $Z$
                & \multicolumn{1}{c}{$\delta^{(4)} g$}
                & \multicolumn{1}{c}{$\Za$-expansion}
 \\
\hline\\[-9pt]
  4 & -3.8x556(1)   &   -3.8x3217 \\
  6 & -3.7x731(1)   &   -3.7x2356 \\
  8 & -3.6x983(1)   &   -3.6x1496 \\
 10 & -3.6x303(1)   &   -3.5x0635 \\
 12 & -3.5x686(1)   &   -3.3x9774 \\
 14 & -3.5x127(1)   &   -3.2x8914 \\
 16 & -3.4x619(1)   &   -3.1x8053 \\
 18 & -3.4x161(1)   &   -3.0x7192 \\
 20 & -3.3x749(1)   &   -2.9x6331 \\
 24 & -3.3x053(1)   &   -2.7x4610 \\
 26 & -3.2x765(1)   &   -2.6x3749 \\
 28 & -3.2x516(1)   &   -2.5x2889 \\
 30 & -3.2x303(1)   &   -2.4x2028 \\
 32 & -3.2x126(1)   &   -2.3x1167 \\
 40 & -3.1x771(1)   &   -1.8x7725 \\
 50 & -3.2x108(1)   &   -1.3x3421 \\
 54 & -3.2x502(1)   &   -1.1x1700 \\
 60 & -3.3x406(1)   &   -0.7x9118 \\
 70 & -3.5x857(1)   &   -0.2x4815 \\
 80 & -3.9x942(1)   &    0.2x9489 \\
 83 & -4.1x575(1)   &    0.4x5780 \\
 90 & -4.6x298(1)   &    0.8x3792 \\
 92 & -4.7x845(1)   &    0.9x4653 \\
100 & -5.6x198(2)   &    1.3x8096 \\
\end{tabular}
\end{ruledtabular}
\end{table}

%
%
\section{Results and discussion}

We now summarize our results obtained for the two-loop QED corrections with closed fermion loops.
Since the previous investigations of these corrections have been performed within the $\Za$
expansion and provided results complete up to order $\alpha^2\,(\Za)^4$, we identify the
higher-order remainder of our numerical results that can be directly added to the results obtained
previously \cite{pachucki:05:gfact}. The higher-order remainders induced by the three sets of
two-loop diagrams with closed fermion loop considered in the present work are summarized in
Table~\ref{tab:ho}. We observe that the S(VP)E diagram yields a very small contribution to the
higher-order remainder, whereas the remainders from the SEVP and VPVP diagrams are large and
comparable in magnitude and enhance each other.

In Table~\ref{tab:gfact}, we collect all presently available contributions for the $1s$
bound-electron $g$ factor for four hydrogen-like ions that are most relevant from the experimental
point of view. For three of them (carbon, oxygen, and silicon), accurate experimental results are
already available \cite{haeffner:00:prl,verdu:04,sturm:13:Si}, whereas for calcium the experiment
is underway \cite{blaum:priv}. Since most of the results collected in Table~\ref{tab:gfact}
appeared previously in the literature, we give here only short comments about the data presented in
the table. The errors of the point-nucleus Dirac value and of the $(\Za)^0$ part of the one-loop
QED correction originate from the uncertainty of the fine-structure constant, $\alpha^{-1} =
137.035\,999\,074(44)$ \cite{mohr:13:codata}. The finite-nuclear-size correction is evaluated with
the standard two-parameter Fermi model of the nuclear charge distribution and the root-mean-square
radii taken from Ref.~\cite{angeli:04}. The error of this correction originates both from the
quoted uncertainty of the rms radius and from the dependence of the result on the model used for
the nuclear-charge distribution.

The results obtained in the present work for the higher-order two-loop corrections with closed
fermion loops are listed in the table under the labels "$\geq$ 2-loop QED, h.o." Since we did not
calculate the complete two-loop QED correction in the present work (the two-loop self-energy
contribution is left out), we do not decrease the overall uncertainty as compared to the previous
investigations. Same as previously \cite{pachucki:05:gfact}, the uncertainty due to higher-order
two-loop contributions is estimated as
\begin{equation}
g^{(2)}_{\rm h.o.} = 2\, g^{(1)}_{\rm h.o.}\,
     \frac{g^{(2)}[(Z\,\alpha)^2]}{g^{(1)}[(Z\,\alpha)^2]}\,,
\end{equation}
where $g^{(n)}_{\rm h.o.}$ is the $n$-loop higher-order QED contribution and
$g^{(n)}[(Z\,\alpha)^2]$ is the $n$-loop $(Z\,\alpha)^2$ QED contribution. We observe that the size
of the two-loop contributions calculated in the present study is about 50\% of the total
uncertainty to the higher-order effects. So, we might refer to the size of the calculated effects
as "expected".

Finally, we comment on the small differences (in the last significant digit) between the data in
Table~\ref{tab:gfact} and the previous compilation in Ref.~\cite{pachucki:05:gfact} for the recoil
corrections. The difference in the first-order ($\sim m/M$) recoil correction is due to the updated
value of the nuclear masses, whereas the difference in the higher-order recoil correction is due to
the updated theoretical result obtained in Ref.~\cite{pachucki:08:recoil} for the arbitrary nuclear
spin.

It is interesting to note that the theoretical prediction for silicon reported in
Table~\ref{tab:gfact} nearly coincides with the experimental result, leaving almost no space for
the higher-order two-loop self-energy correction, which remains uncalculated at present. Indeed,
the difference between theoretical and experimental results for Si amounts to $4\times 10^{-10}$,
which is twice smaller than the two-loop contribution calculated in the present work. This
indicates that either the two-loop self-energy contribution is relatively small or it changes its
sign in the vicinity of $Z = 14$. In order to test these assumptions, a $g$-factor measurement in a
heavier system would be of great help. Table~\ref{tab:gfact} shows that already for calcium, the
uncertainty due to the two-loop self-energy is by two orders of magnitude larger than the other
theoretical errors. So, a measurement of the bound-electron $g$ factor in Ca$^{19+}$ with the same
accuracy as in Si would lead to an unambiguous experimental determination of the higher-order
two-loop self-energy contribution.

Summarizing, we have calculated three sets of two-loop QED diagrams with the closed fermion loops
to the $1s$ bound-electron $g$ factor. Calculations were performed to all orders in the nuclear
binding strength parameter $\Za$ except for the closed fermion loop, which was treated within the
free-loop (Uehling) approximation in some cases. Our numerical data were shown to agree well with
the $\Za$-expansion results previously obtained for these corrections. The higher-order remainder
[of order $\alpha^2(\Za)^5$ and higher] was separated out from our numerical results. Its size
agrees well with previous estimations for the two-loop higher-order effects. Our calculations do
not improve the total uncertainty of the two-loop QED effects in the theoretical predictions since
the most nontrivial two-loop correction, the two-loop self-energy, still remains to be calculated.

\section*{Acknowledgement}

The work presented in the paper was supported by the Alliance Program of the Helmholtz Association
(HA216/EMMI).

\begin{table*}
\setlength{\LTcapwidth}{\columnwidth} \caption{The higher-order contribution to the $1s$
bound-electron $g$ factor induced by the two-loop QED diagrams with closed fermion loops, in terms
of $\delta^{(5)} g = \Delta g/[(\alpha^2/\pi^2)(\Za)^5]$, where $\Delta g$ is the contribution to
the $g$ factor.
 \label{tab:ho} }
\begin{ruledtabular}
\begin{tabular}{l.....}
                $Z$
                & \multicolumn{1}{c}{S(VP)E}
                & \multicolumn{1}{c}{SEVP}
                & \multicolumn{1}{c}{VPVP}
                                         & \multicolumn{1}{c}{$\delta^{(5)} g$}
                                         & \multicolumn{1}{c}{$\Delta g \times 10^6$} \\
\hline\\[-9pt]
  6 &     0.x00(15)^a &     7.x97(36)   &     6.x31       &    14.x28(39)   &          0.000x\,012\,4(3)   \\
  8 &    -0.x05(12)^a &     8.x03(6)    &     6.x01       &    14.x00(13)   &          0.000x\,051\,2(5)   \\
 14 &    -0.x15(4)^a  &     7.x62(1)    &     5.x25       &    12.x72(4)    &          0.000x\,764\,(3)   \\
 20 &    -0.x23(2)^a  &     7.x15(1)    &     4.x62       &    11.x54(2)    &          0.004x\,12(1)   \\
 30 &    -0.x27(2)    &     6.x56       &     3.x74       &    10.x03(2)    &          0.027x\,2(1)   \\
 40 &    -0.x22       &     6.x19       &     2.x99       &     8.x96       &          0.102x\,4(1)   \\
 50 &    -0.x12       &     6.x03       &     2.x30       &     8.x21       &          0.286x\,5(1)   \\
 60 &     0.x01       &     6.x07       &     1.x62       &     7.x70       &          0.668x\,2(1)   \\
 70 &     0.x16       &     6.x30       &     0.x91       &     7.x37       &          1.382x\,3(2)  \\
 80 &     0.x33       &     6.x78       &     0.x09       &     7.x20       &          2.632x\,7(3)   \\
 83 &     0.x38       &     6.x97       &    -0.x18       &     7.x17       &          3.155x\,0(3)   \\
 90 &     0.x53       &     7.x52       &    -0.x88       &     7.x17       &          4.726x\,(1)   \\
 92 &     0.x58       &     7.x69       &    -1.x10       &     7.x18       &          5.281x\,(1)   \\
100 &     0.x81       &     8.x63       &    -2.x15       &     7.x28       &          8.134x\,(3)   \\
\end{tabular}
\end{ruledtabular}
$^a$\ extrapolated value.
\end{table*}

%
%
\begingroup
\squeezetable
\begin{table*}[htb]
\begin{center}
\caption{Individual contributions to the $1s$ bound-electron $g$ factor. The abbreviations used are
as follows: ``h.o.'' stands for a higher-order contribution, ``VP-EL'' -- for the electric-loop
vacuum-polarization correction, ``VP-ML'' -- for the magnetic-loop vacuum-polarization correction,
``TW'' indicates the results obtained in this work. $\langle r^2\rangle^{1/2}$ is the
root-mean-square nuclear charge radius in fermi. \label{tab:gfact} }
\begin{ruledtabular}
\begin{tabular}{ll....c}
& & \multicolumn{1}{c}{$^{12}{\rm C}^{5+}$ } & \multicolumn{1}{c}{$^{16}{\rm O}^{7+}$} &
\multicolumn{1}{c}{$^{28}{\rm Si}^{13+}$} & \multicolumn{1}{c}{$^{40}{\rm Ca}^{19+}$}
& Ref. \\
$\langle r^2\rangle^{1/2}$ &           & \multicolumn{1}{c}{2.4703\,(22)}  &  \multicolumn{1}{c}{2.7013\,(55)} & \multicolumn{1}{c}{3.1223\,(24)} & \multicolumn{1}{c}{3.4764\,(10)} & \\
$m/M$                          && \multicolumn{1}{c}{$4.5728 \times 10^{-5}$} & \multicolumn{1}{c}{$3.4307 \times 10^{-5}$}& \multicolumn{1}{c}{$1.9614 \times 10^{-5}$}& \multicolumn{1}{c}{$1.3731 \times
10^{-5}$}\\\hline\\[-0.15cm]
\multicolumn{2}{l}{Dirac value (point)}&  1.998\, 721x\, 354\, 39\,(1)    & 1.997\, 726x\, 003\, 06\,(2)   &   1.993\,023x\,571\,6      & 1.985\, 723x\, 203\, 7\,(1)  & \\
Finite nuclear size  &                 &  0.000\, 000x\, 000\, 41         & 0.000\, 000x\, 001\, 55\,(1)   &   0.000\,000x\,020\,5      & 0.000\, 000x\, 113\, 0\,(1)  & \\
1-loop QED      & $     (Z\,\alpha)^0$ &  0.002\, 322x\, 819\, 47\,(1)    & 0.002\, 322x\, 819\, 47\,(1)   &   0.002\,322x\,819\,5      & 0.002\, 322x\, 819\, 5       &    \\
                & $     (Z\,\alpha)^2$ &  0.000\, 000x\, 742\, 16         & 0.000\, 001x\, 319\, 40        &   0.000\,004x\,040\,6      & 0.000\, 008x\, 246\, 2       & \cite{grotch:70:prl}\\
                & $     (Z\,\alpha)^4$ &  0.000\, 000x\, 093\, 42         & 0.000\, 000x\, 240\, 07        &   0.000\,001x\,244\,6      & 0.000\, 002x\, 510\, 6       & \cite{pachucki:04:prl} \\
                & h.o., SE             &  0.000\, 000x\, 008\, 28         & 0.000\, 000x\, 034\, 43\,(1)   &   0.000\,000x\,542\,8\,(3) & 0.000\, 003x\, 107\, 7\,(2)  & \cite{yerokhin:02:prl,pachucki:04:prl} \\
                & h.o., VP-EL          &  0.000\, 000x\, 000\, 56         & 0.000\, 000x\, 002\, 24        &   0.000\,000x\,032\,6      & 0.000\, 000x\, 172\, 7       & \cite{beier:00:rep} \\
                & h.o., VP-ML          &  0.000\, 000x\, 000\, 04         & 0.000\, 000x\, 000\, 16        &   0.000\,000x\,002\,5      & 0.000\, 000x\, 014\, 6       & \cite{lee:05} \\
$\ge$2-loop QED &$     (Z\,\alpha)^0$  & -0.000\, 003x\, 515\, 10         &-0.000\, 003x\, 515\, 10        &  -0.000\,003x\,515\,1      & -0.000\, 003x\, 515\, 1       & \\
                &$     (Z\,\alpha)^2$  & -0.000\, 000x\, 001\, 12         &-0.000\, 000x\, 002\, 00        &  -0.000\,000x\,006\,1      & -0.000\, 000x\, 012\, 5       & \cite{grotch:70:prl}\\
                &$     (Z\,\alpha)^4$  &  0.000\, 000x\, 000\, 06         & 0.000\, 000x\, 000\, 08        &  -0.000\,000x\,001\,3      & -0.000\, 000x\, 010\, 9       & \cite{pachucki:05:gfact} \\
                & h.o.                 &  0.000\, 000x\, 000\, 01\,(3)    & 0.000\, 000x\, 000\, 05\,(11)  &   0.000\,000x\,000\,8\,(17)& 0.000\, 000x\, 004\, 1\,(100)&  TW \\
Recoil          &$ m/M$                &  0.000\, 000x\, 087\, 73         & 0.000\, 000x\, 117\, 10        &   0.000\,000x\,206\,1      & 0.000\, 000x\, 297\, 4       & \cite{shabaev:02:recprl} \\
                & h.o.                 & -0.000\, 000x\, 000\, 10         &-0.000\, 000x\, 000\, 13        &  -0.000\,000x\,000\,2      & -0.000\, 000x\, 000\, 3       & \cite{pachucki:08:recoil}  \\[0.15cm]
Total           &                      &  2.001\, 041x\, 590\, 20\,(3)    & 2.000\, 047x\, 020\, 38\,(11)  &   1.995\,348x\,958\,7\,(17)&1.988\, 056x\, 950\, 7\,(100)  \\
Experiment \cite{haeffner:00:prl,verdu:04,sturm:13:Si}
                &                      &  2.001\, 041x\, 596\,(5)         & 2.000\, 047x\, 025\,4\,(46)    &   1.995\,348x\,959\,10\,(81)\\
\end{tabular}
\end{ruledtabular}
\end{center}
\end{table*}
\endgroup


\appendix

\widetext

\section{S(VP)E correction: zero-potential contribution}
\label{app:1}

The zero-potential contribution of the S(VP)E correction can be derived by the method described in
Ref.~\cite{yerokhin:04} (Sec.~IIIA) for the one-loop SE correction to the $g$ factor. Derivation requires
explicit expressions for the free SE and vertex operators with the VP insertion into the photon line,
which were obtained previously in Ref.~\cite{yerokhin:08:pra} (Sec.~IIIC). The total zero-potential
contribution of the S(VP)E correction to the $g$ factor is separated into three parts, which have
the same meaning as in Ref.~\cite{yerokhin:04},
\begin{equation}
\Delta g_{\rm vr}^{(0)} = \Delta g_{\rm ver,1}^{(0)} + \Delta g_{\rm ver,2}^{(0)} + \Delta g_{\rm
red}^{(0)}\,.
\end{equation}
The results for the three terms in the above equation are:
\begin{align}\label{ap1}
\Delta g_{\rm ver,1}^{(0)} & \ = -\frac{\alpha^2}{4\pi^5}\,\int_0^{\infty}p_r^2dp_r\,
 \int_0^1dx\,dz\, \frac{z^2(1-z^2/3)}{1-z^2}\, \frac{1-x}{\Delta(\rho)}\,
\biggl[ g_a(\vare_a g_a+ p_r\, f_a) -\frac13 f_a(\vare_a f_a + p_r\, g_a)\biggr]\,,
\end{align}
\begin{align}
\Delta g_{\rm ver,2}^{(0)} & \ = \frac{\alpha^2}{12\pi^5}\,\int_0^{\infty}p_r^2dp_r\,
 \int_0^1dx\,dz\, \frac{z^2(1-z^2/3)}{1-z^2}\, \frac{1-x}{\Delta(\rho)}\,
  \nonumber \\ &  \times
\biggl\{\Delta(\rho)\,\ln \frac{\Delta(\rho)}{\Delta(0)}\biggl[ \frac2{p_r}\,g_af_a + g_af_a^{\prime}
+ f_ag_a^{\prime}\biggr]
+2(1-x) (\vare_af_a+ p_r\,g_a)f_a - 4 f_a^2 \biggr\}\,,
\end{align}
\begin{align}\label{ap3}
\Delta g_{\rm red}^{(0)} & \ = g_D\, \frac{\alpha^2}{16\pi^5}\,\int_0^{\infty}p_r^2dp_r\,
 \int_0^1dx\,dz\, \frac{z^2(1-z^2/3)}{1-z^2}\, \frac{1-x}{\Delta(\rho)}\,
  \nonumber \\ &  \times
\biggl\{\Delta(\rho)\,\ln \frac{\Delta(\rho)}{\Delta(0)} (g_a^2 + f_a^2)
-2\vare_a (1-x) \biggl[ \vare_a(g_a^2+f_a^2)+2p_r\,g_af_a\biggr]
+ 4\vare_a\,(g_a^2-f_a^2)\biggr\}\,,
\end{align}
where $\Delta(\rho) = x(1-\rho)+\rho+ 4(1-x)/[x(1-z^2)]$, $\rho = (m^2-p^2)/m^2 = (m^2-\vare_a^2+p_r^2)/m^2$,
$g_D$ is the lowest-order (Dirac) bound-electron $g$ factor, which for the $1s$ state is given by
\begin{align}
g_D = \frac23\left( 1 + 2\,\sqrt{1-(Z\alpha)^2}\right)\,,
\end{align}
$\vare_a$ is the reference-state energy, $g_a \equiv g_a(p_r)$ and $f_a \equiv f_a(p_r)$ are the
upper and the lower components of the reference-state wave function in the momentum representation,
respectively, and $g_a^{\prime}$ and $f_a^{\prime}$ denote derivatives of $g_a(p_r)$ and $f_a(p_r)$
with respect to $p_r$.

\end{document}